\documentclass[sigconf,table]{acmart} % anonymous

\usepackage{graphicx}
\usepackage{enumitem}
\usepackage{tabularx}
\usepackage{makecell}
\usepackage{booktabs}
\usepackage{multirow}
\usepackage{xcolor}

\graphicspath{{img/}}

\newcommand{\subscript}[2]{$#1 _ #2$}
\newcommand{\paragraphb}[1]{\vspace{0.03in}\noindent{\bf #1} }

\begin{document}

\title{Reconstructing Robot Operations via Radio-Frequency Side-Channel}

\author{Ryan Shah}
%\institute{University of Strathclyde, Glasgow}
\affiliation{%
	\institution{University of Strathclyde}
	% \city{Glasgow}
	\country{United Kingdom}
}
\email{ryan.shah@strath.ac.uk}

\author{Mujeeb Ahmed}
%\institute{University of Strathclyde, Glasgow}
\affiliation{%
	\institution{University of Strathclyde}
	% \city{Glasgow}
	\country{United Kingdom}
}
\email{mujeeb.ahmed@strath.ac.uk}

\author{Shishir Nagaraja}
%\institute{University of Strathclyde, Glasgow}
\affiliation{%
	\institution{Newcastle University}
	% \city{Newcastle}
	\country{United Kingdom}
}
\email{shishir.nagaraja@newcastle.ac.uk}

\begin{abstract}
Connected teleoperated robotic systems play a key role in ensuring operational workflows are carried out with high levels of accuracy and low margins of error. In recent years, a variety of attacks have been proposed that actively target the robot itself from the cyber domain. However, little attention has been paid to the capabilities of a passive attacker. In this work, we investigate whether an insider adversary can accurately fingerprint robot movements and operational warehousing workflows via the radio frequency side channel in a stealthy manner. Using an SVM for classification, we found that an adversary can fingerprint individual robot movements with at least 96\% accuracy, increasing to near perfect accuracy when reconstructing entire warehousing workflows.
\end{abstract}



\maketitle

% Introduction
\section{Introduction}
\label{sec:intro}

Teleoperated robotics systems are becoming more prominent in many applications,
in particular surgical~\cite{Tewari2002,Hannaford2012} and industrial environments~\cite{kumar2018development,grabowski2021teleoperated,bartovs2021overview}. %\textcolor{red}{CITATION} 
The key reason for their
employment pertains to promises of higher efficiency and accuracy, and lower
margins of error, compared to a human traditionally operating in the same
manner. Many of these robotic systems are Internet-connected, which leaves them
vulnerable to attack and compromise in the cyber domain~\cite{Bonaci2015,McClean2013,DeMarinis2019}. %\textcolor{red}{CITATION}
These attacks vary
from manipulating messages and feedback to or from the operator, to directly
hijacking the robot's controller directly. However, most of this research
focuses on active attackers, with little focus given to reconnaissance aspects
in a passive manner. While active attacks can be deadly to the operating
environment and subject(s) involved, passive attacks can result in huge losses
that stem from stealthy, unintentional information leakage. For example, if
an attacker is able to identify what workflows a robot is carrying out, such as the movement of packages in a warehouse between belts, they could use this
information to sell on to competitors that can understand how competing
warehousing facilities operate and use this information to a malicious
advantage~\cite{maggi2017rogue,pogliani2019security}. %\textcolor{red}{Can we point to some news article or research about this kind of threat/attacks?}

In this work we seek to explore other mechanisms to passively learn about robotic workflows. Side channels have previously been used in different technological domains as a means to learn sensitive information about the devices, processes and algorithms~\cite{song2016my,chhetri2017confidentiality,sami2020spying}. One particularly interesting side channel looks at IP theft via the power side channel in 3D printers~\cite{gatlin2021encryption}, focusing on predictable movements and reconstructing G-code corresponding to IP. In the case of many robots, however, movement inference is much more dynamic and unpredictable. In this work, we propose a side channel technique to mount an information leakage attack, leveraging unintentional radio frequency (RF) emissions during normal robot operations. While such information could be used to compromise IP, information leakages from robotic workflows as previously described could lead to wholesome compromise of operational confidentiality of organisations.  %\textcolor{red}{CITATIONS}.
In a typical industrial robot, for example, components such as stepper motors and microprocessors are used which can emanate unintentional radio frequencies~\cite{cobb2010physical}. By conducting a feasibility test which explored whether our robot did in fact emit unintentional radio frequencies, given the use of the same components, we found that this was in fact true.
% \textcolor{red}{Say something about the side channel relationship and the workflow here...3-4 lines....for example, robotic movements are controlled by motors and those motors produce EM emanations but give few more sentences in this respect befoe going in to details of what you did}
We collected samples of RF signals
corresponding to both individual robot movements along each of three axes (X, Y and Z) and their permutations, as well as those corresponding to entire warehousing workflows (i.e. picking packages from one location and
placing them in another). We then follow a inference technique to extract
a small subset of features that represent enough variance to construct an
accurate movement or workflow fingerprint. Through the use of an SVM classifier,
we observe at least 96\% accuracy as a baseline set of results, with similar
results observed for more fine-grained information such as how fast or long the
robot moves. Further, we show that entire workflows can be reconstructed with
even higher accuracy than a pattern-matching approach from individual movements
alone.

\paragraphb{Organisation.} This paper is organised as follows.
In Section~\ref{sec:background}, we provide background information on radio
frequency and teleoperated robotic systems, and set out the threat model for
this attack. We then discuss our attack methodology and evaluate the attack in
Section~\ref{sec:methodology}. A discussion on our findings and insights is
found in Section~\ref{sec:discussion} with related work provided in
Section~\ref{sec:related}. Finally, we conclude in Section~\ref{sec:conclusion}.

% Background
\section{Background}
\label{sec:background}

\subsection{Radio Frequency}

Radio Frequency (RF) is used in a wide array of application contexts, ranging
from TV and radio, to wireless and satellite communications. Specifically, it
is a measurement which represents the oscillation rate of EMF waves whose
frequencies are within range of \textasciitilde3kHz--300GHz, and alternating
currents that carry the signal. While the term radio frequency refers to the
range of 3kHz--300GHz, this is actually the radio spectrum in which several
frequency bands are defined in which different transmission systems operate.
For example, AM radio operates within 600kHz--1.6MHz whilst audible sound waves usually broadcast between 20Hz--10kHz.

There are two oscillation types, electrical and mechanical oscillation, however
most cases of RF refer to electrical oscillations. Electronic circuits emit some
degree of electromagnetic emissions while they operate. While these emissions
can theoretically span the entire spectrum, the focus of this study is on
emissions that fall within the radio frequency spectrum. While there may be
many significant sources of unintentional RF emissions, many robotic systems
will make use of microprocessors and stepper motors. The former has been
demonstrated to emit RF due to switching activities of transistors alternating
varying current flows~\cite{cobb2010physical}. Similarly, in the case of the
stepper motors, digital pulses and phase shifts in voltage may also contribute
to RF emissions.

\subsection{Teleoperated Industrial Robotic Systems}

Teleoperated robotic systems play pivotal roles in safety-critical environments such as surgical theatres and manufacturing facilities, providing high levels of accuracy and precision, and lower margins of error compared to humans performing the same task(s). In industrial settings, notable examples include the likes of robotic arms (and others) in environments such as in production lines~\cite{aschenbrenner2015teleoperation} and product warehouses~\cite{avila2020study,grabowski2021teleoperated}, among others~\cite{goto2010teleoperation,li2017teleoperation,toquica2019web,kamali2020real}.

\begin{figure}[h]
	\centering
	\includegraphics[width=1.0\linewidth]{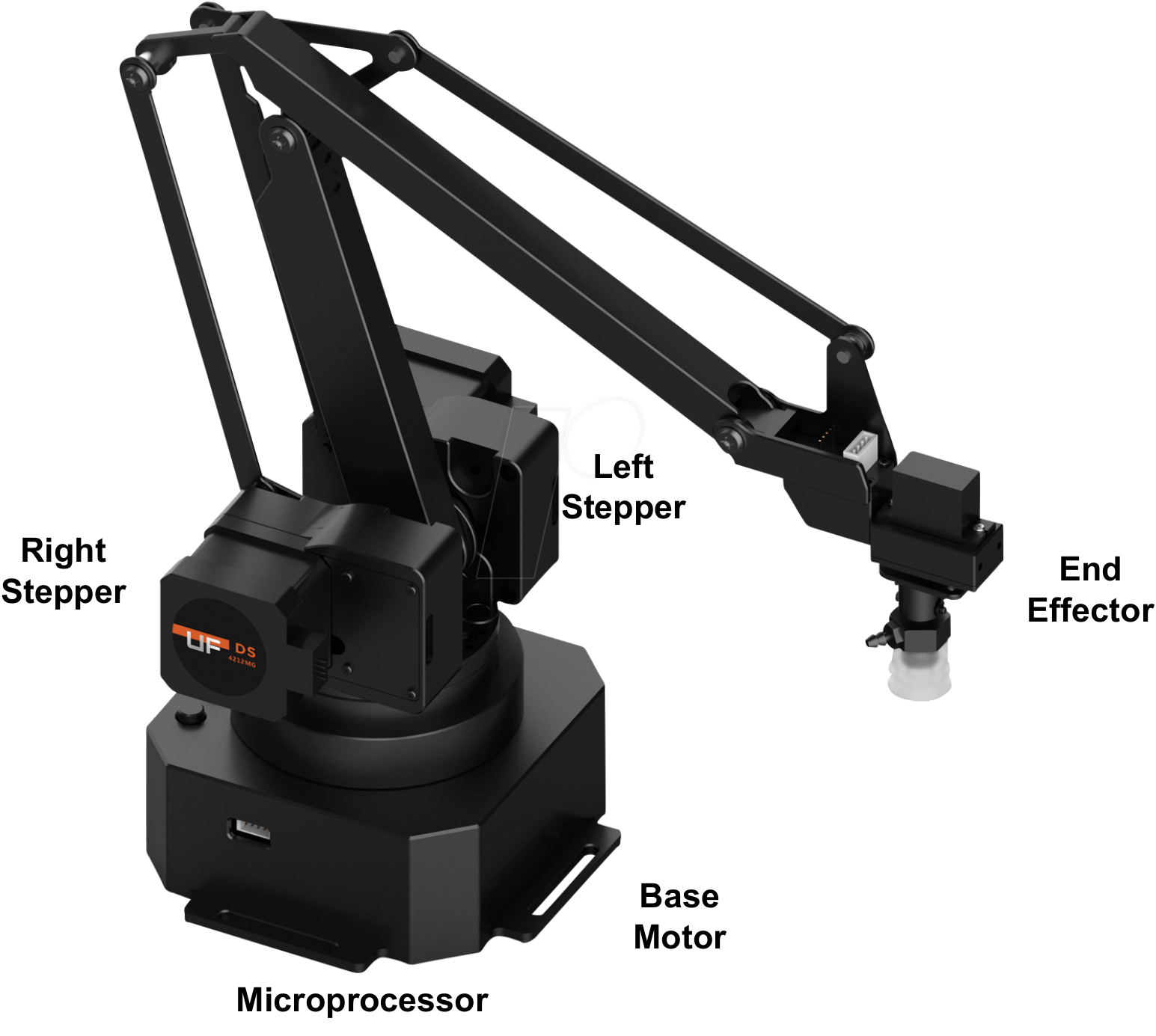}
	\caption{Robot Components}
	\label{fig:robotcomponents}
\end{figure}

While such systems span a variety of environments, they all share a common system architecture~\cite{quarta2017experimental}. Simply, there exists: a controller (i.e. a teach pendant) which has physical controls or a touch-screen interactive GUI to control the robot; a set of input and output devices (sensors and actuators); and a network in which the robot operates, all linked together via an electronic control system. The primary link in these architectures is that between the controller and electronic control system, and is considered safety-critical due to the ability to directly influence the actions the robot can and will perform. In this paper, we make use of a robotic arm -- common to many industrial applications -- which consists of three stepper motors for each of the three axes (Figure~\ref{fig:robotcomponents}). The base motor controls the Y axis while the left and right control the X and Z axes. In some cases, the microprocessor may be located in another physical component between the robot and controller, however in others it may be situated in the robot itself. In this robot, the microprocessor is located in the base structure alongside the stepper motor. We believe that this robot is a suitable replication candidate for this attack as it compares to other single-arm robots used in typical industrial settings such as warehouses, which have at least 3 degrees-of-freedom, and also consists of the same principal components for movement fingerprinting (microprocessor
and stepper motors). With regard to potential emanations of unintentional RF, the primary component of interest is the microprocessor in the base of the arm robot. However, given that stepper motors may also contribute to sources of RF, these are also of interest to an attacker.
% \textcolor{red}{Write 1-2 paragraphs explaining your particular robot....saying this is used in this study....also if possible paste a construction diagram of the robotic arm showing major components, including motors, controller etc... and explain the sources of EM emanation...I know in Sec 3.1 you mention name of the robotic arm etc but that is a good fit for there and keep it there....What I am asking here is slightly different...Robot constrcution with an explanation on what constittutes the kind of side channel that we are looking at....this will convinve the reader on the validity of your experiements }

\begin{figure*}[h]
	\centering
	\includegraphics[width=0.8\linewidth]{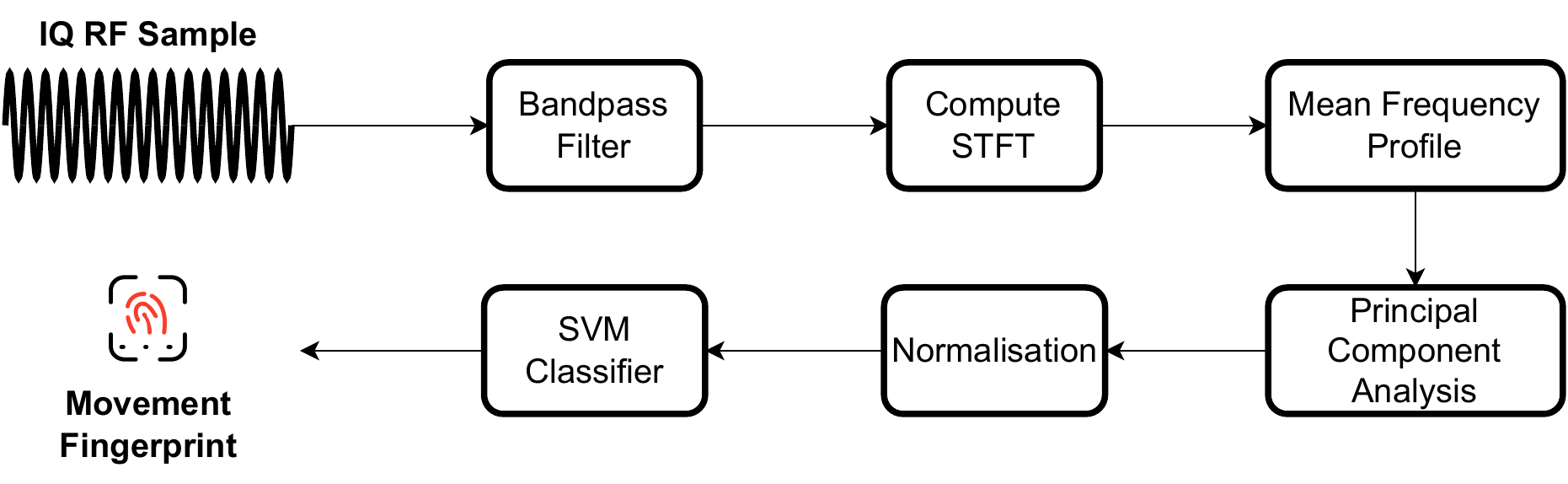}
	\caption{Attack Methodology} %\textcolor{red}{Need to improve the figure quality...consider using both columns for this figure....bigger font size and bigger width of lines....It is totally not visible in print version and without zoom in digital.... } }
	\label{fig:attackmethod}
\end{figure*}

\subsection{Threat Model}

In a teleoperated architecture, many attacks have been presented in literature.
These range from tampering with control commands, modifying feedback, and
physical-domain attacks (i.e. modifying operational environment), among others~\cite{McClean2013,Bonaci2015}. %\textcolor{red}{CITATION} .
However, most of these attacks focus on an active adversary with little
attention paid to passive (weaker) adversaries. In this work, we look at the
capabilities of a passive adversary.

The goal in this context is a {\em stealthy} approach to fingerprint the
movements a teleoperated industrial robot can carry out via the radio-frequency
side channel. If successful, robot movement fingerprints can be used to not
only determine possible workflows, but also be used to reconstruct these
workflows. In this work, we look at the context of a logistics warehouse where
products are packaged and stored or moved around the warehouses (i.e. along
conveyor belts) to then progress to the next stage in a supply chain.

The primary adversary we consider in the scope of this study is an insider, such
as technical staff who operate or maintain the robot on the warehouse floor. To
capture unintentional RF emissions, an RF receiver is used. These have varying
ranges for frequency capture but also can vary in physical size as well. A
passive insider here should be stealthy to avoid possible detection, and thus
the question pertaining to this is, does there exist a small enough receiver to
be used to capture the RF emanations from the robot to minimise the possibility
of detection. Aside from the ability to conduct the attack, and other important
consideration relates to the opportunities that arise from the successful
collection of movement fingerprints. By collecting movement fingrprints, it
would be possible to correlate a series of movements with known patterns that
correspond to warehousing workflows such as picking and placing products between
two locations. Furthermore, it may also be possible to reconstruct workflows
directly, by capturing the unintentional RF emanations at the time of these
workflows and later performing direct workflow classification (as opposed to
workflow identification from a collection of movement series). In either case,
the information leakage of these operational movements can expose a degree of
operational-level detail that may not be otherwise available to this attacker.
First, operational confidentiality in terms of claimed efficiency of warehousing
procedures or what operations are carried out in the warehouse could be collected
and be used as bribery, leaked to competitors, or be used to discredit an
organisation at the expense of compromising operational confidentiality. Third,
it may also be possible to identify the contents of packages. While gripping or
lifting actions alone may not contribute to an increase in unintentional RF
emissions, the weight of products may in fact do so. Specifically, heavier
weights inherently require more force to be exerted on the robot to lift products
and thus require more power to successfully do so, resulting in increased RF
emission.

\subsubsection{Hypotheses and Goals}
\label{sec:goals}

In this work, we aim to investigate whether it is indeed possible to passively
fingrprint teleoperated robot movements via the RF side channel. While we
hypothesise that this will be possible, we question the extents to which this
is considered a successful attack. Thus, we propose the following research
questions:

\begin{enumerate}[label=(\subscript{R}{{\arabic*}})]
	\item Can an adversary identify teleoperated robot movements via unintentional
		  RF emissions?
	\item Can these robot movements be fingerprinted with more granularity?
		  Specifically, does the speed or distance of movement impact the
		  accuracy of fingerprinting?
	\item Can {\em higher-level} warehousing workflows be reconstructed from
		  unintentional RF emissions?
	% \item Can more granularity for information leakage of workflows be attained
	% 	  by observing variations in products being interacted with (i.e. weight)?
	\item Can this attack be performed stealthily, using an RF antenna/receiver
		  that can potentially operate unnoticed?
	% \item How can an organisation defend against this attack?
\end{enumerate}

% Attack
\section{Attack Methodology}
\label{sec:methodology}

As previously stated, the goals of this study are to: (a) fingerprint individual
movements of an industrial robot arm, and (b) reconstruct warehousing
workflows, using the RF side channel (unintentional RF emissions). An overview
of our attack methodology can be seen in Figure~\ref{fig:attackmethod}. We
first apply a Butterworth bandpass filter to eliminate undesired frequency content. We
then compute the Short-Time Fourier Transform (STFT) to represent the frequency content
over time and compute the Mean Frequency Profile (MFP) to observe width to frequency peaks and
the cadence of robot movements. We then apply Principal Component Analysis (PCA) for
dimensionality reduction and normalise the feature sets before fingerprinting.
Further detail to each of the steps in our methodology are subsequently
described below. % \textcolor{red}{In 2-3 lines please summarise all the blocks in Fig2 here and then say that the detailed description is in the following. This will serve as a summary and following content discusses each aspect in further details}

\subsection{Experimental Setup}

\subsubsection{Robot Environment}
In the first stage of our setup, we discuss our robotics environment. The focus of this study is on teleoperated industrial robots, particularly those used in warehouses. In this architecture, a robot (consisting of sensors, actuators, etc.) is paired with a controller (i.e teach pendant) that is used to send commands or execute pre-programmed actions which the robot can interpret and execute on the factory floor. In our study, we replicate this on a smaller scale, we used uFactory's uARM Swift Pro%~\cite{ufactory}
operated by an Arduino Mega 2560
running MicroPython. The controller is run on a Windows 10 laptop running
the uARM Python (3.8.X) SDK%~\cite{uarmdeveloper}.
% \textcolor{red}{This is right place to put a picture of real experiment setup properly labelled and refer it here...as you are explaining each component like controller, antenna etc...it would help reviewers when they see the real setup...they highly appreciate that....}

\begin{figure}[ht]
	\centering
	\includegraphics[width=1.0\linewidth]{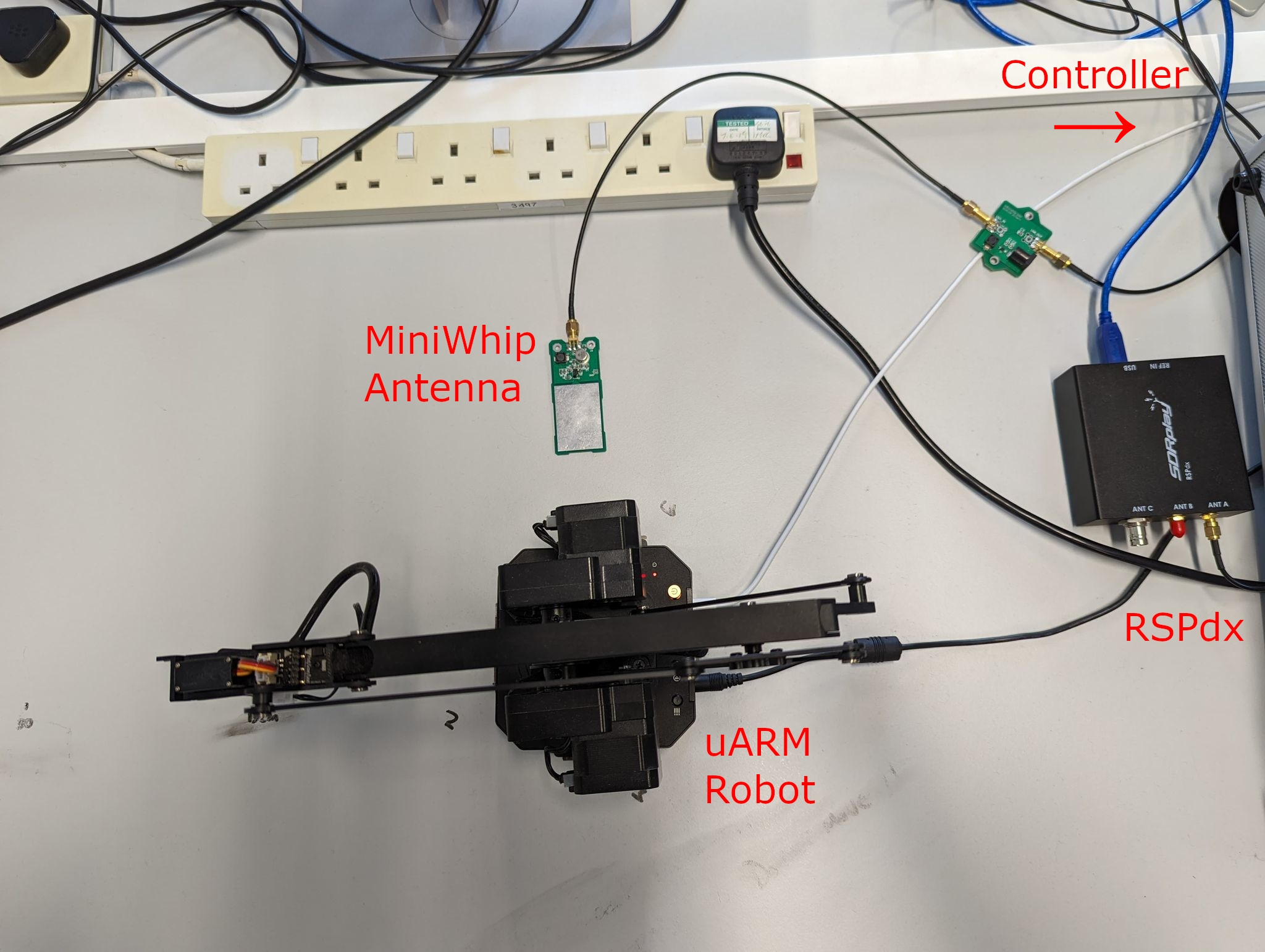}
    \captionsetup{singlelinecheck=off}
    \caption{\centering Robot Environment\hspace{\textwidth}{\textcolor{darkgray}{\small\textmd{The attack
    environment consists of the robot, MiniWhip antenna, RSPdx SDR and path to controller}}}}
	\label{fig:robotenv}
\end{figure}

To capture RF emissions, we used a Mini-Whip Medium-Shortwave Active Antenna
which is placed near the base of the robot arm. The antenna amplifies the
unintentional RF signal (emitted during robot operations) and transmits this to an RTL-SDR receiver via a shielded coaxial cable. This antenna is suitable for the scale of our robot in this study, with an operating frequency of
10kHz--30MHz and a small physical size of $113*32*7mm$. The small physical size aims to demonstrate that this attack can be conducted stealthily and {\em hide} by the robot ($R_4$). The RF emissions were captured as IQ files using SDRuno %~\cite{sdrplay2022}
 at a sampling rate of 2MHz.
An overview of the robot and antenna setup can be seen in Figure~\ref{fig:robotenv}, including the robot, SDRplay, antenna and pointer towards the controller.

\subsubsection{Movement Dataset}
\label{sec:dataset}

\begin{figure}
	\centering
	\includegraphics[width=1.0\linewidth]{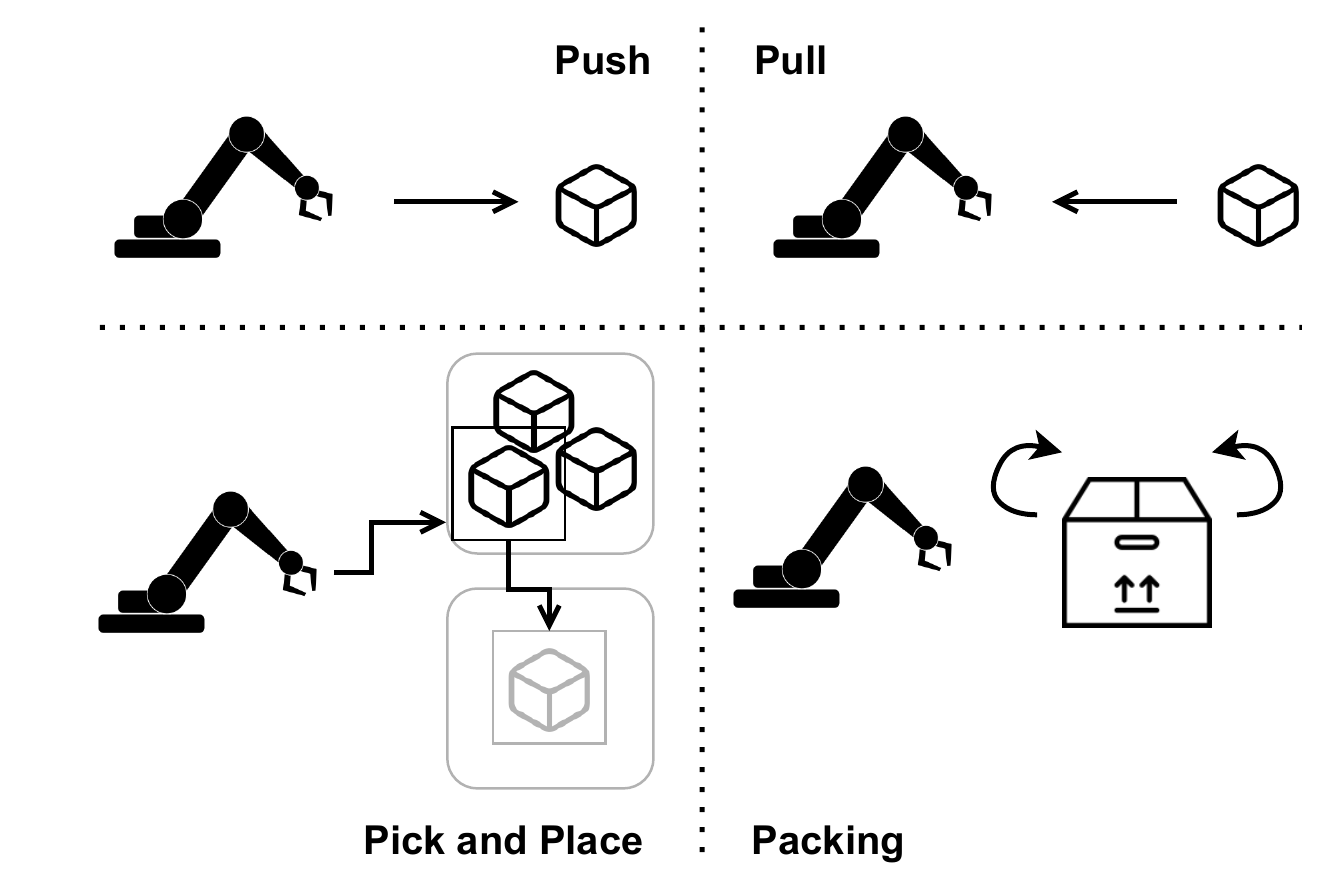}
    \captionsetup{singlelinecheck=off}
    \caption{\centering Depiction of Common Warehousing Workflows\hspace{\textwidth}{\textcolor{darkgray}{\small\textmd{Our dataset contains common warehousing workflows such as pushing, pulling, packing and moving objects}}}}
	\label{fig:workflows}
\end{figure}

The next stage of our experimental setup was to create our movement dataset.
In this dataset, we have sets of samples which correlate with our objectives
and thus, is split into two subsets. The first subset contains samples
pertaining to all permutations of X, Y, and Z with varying speeds of movement
(12.5--100mm/s) and distances (1--50mm) ($R_1-R_2$). The second contains samples of
manufacturing workflows ($R_3$) such as pick-and-place, push and pull operations which
were replicated from those found in existing industrial robot datasets such as
the {\em Forward Dynamics Dataset Using KUKA LWR and
Baxter}~\cite{polydoros2016reservoir} for pick and place and the {\em Inverse
Dynamics Dataset Using KUKA}~\cite{rueckert2017learning} for push/pull. A
depiction of these workflows can be seen in Figure~\ref{fig:workflows}. We
observe that the core information which details these workflows are the dynamic
movements being carried out, which are potentially influenced by additional
input (i.e. from sensors). As well as this, we also collect samples for
pick-and-place operations with packages of different weights (up to maximum
load of 1kg for the robot arm), to investigate if content weights can be
detected via RF side-channel. Further, also within this second subset, we also
perturbed existing data to form additional samples to account for a small
degree of entropy that may be present in real-world operations (i.e. those
that may arise due to drift in equipment calibration or wear-and-tear).
As a whole, the first subset contains \textasciitilde7.8K samples for individual
movements and the second containing \textasciitilde400 samples for warehousing
workflows, with each using 20\% of total samples for testing.

\subsection{Feature Extraction and Movement Fingerprinting}

After establishing our movement dataset, the next step in our attack methodology
was to extract features from each of the signal samples which would later be
used for fingerprinting. The goal here is to ensure that each signal sample for a movement can be easily distinguished from other movements.

\begin{figure*}[h]
	\centering
	\includegraphics[width=1.0\linewidth]{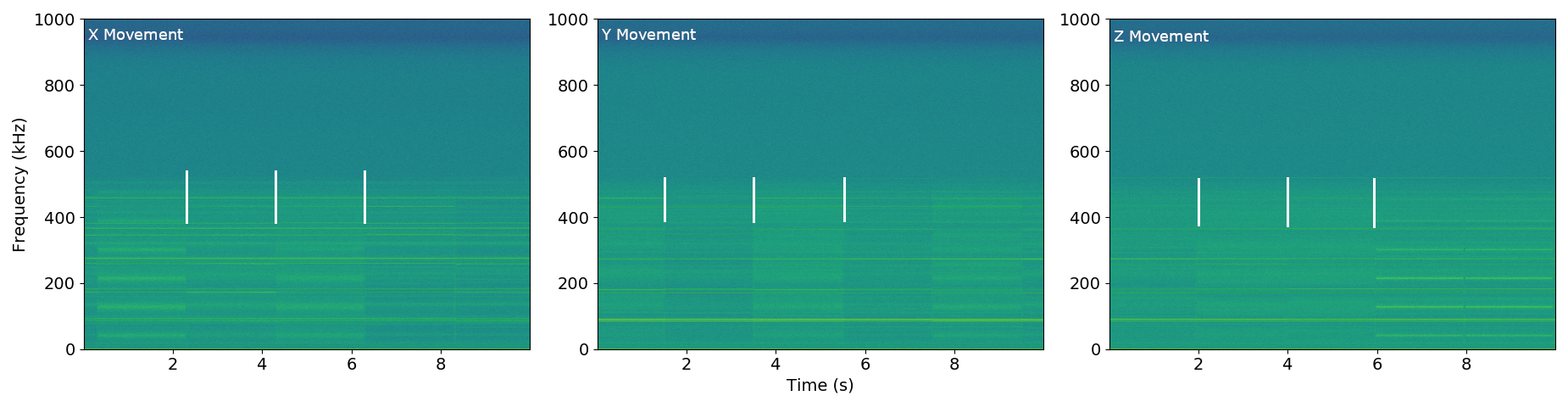}
    \captionsetup{singlelinecheck=off}
    \caption{\centering Observing Frequency Peaks for X, Y and Z Movements\hspace{\textwidth}{\textcolor{darkgray}{\small\textmd{Observing log-spectra for movements shows movement peaks and demonstrates a visual approach may not be suitable for distinguishing variance within and between movements}}}}
	\label{fig:peaks}
\end{figure*}

First, we wanted to observe whether we could see peaks in the captured RF
emissions corresponding to robot movements. To do this, we take the Short-Time
Fourier Transform (STFT) of a set of signal samples and compute the log-spectra
(spectrogram) using a segment length of 8192 and a Hann window as a default,
first observation. The STFT allows us to observe information which concerns
variations of frequency content of the signal over time. For this, we programmed
the robot to perform an X movement at 2s intervals over a period of 10 seconds.
As illustrated in Figure~\ref{fig:peaks}, we can indeed observe peaks at 2s
intervals, which correspond with the robot moving as programmed. However, it is
clear that variations present within and between movements are not easily
distinguishable through a visual approach. From this, we carry out feature
extraction for each sample which will be used as input for movement
fingerprinting. The techniques applied for feature extraction are a variation
of those presented in the work by Zabalza et al.~\cite{zabalza2014robust}
which demonstrates good classification accuracy for identifying moving targets
via radar systems.

\begin{figure}[H]
	\centering
	\includegraphics[width=1.0\linewidth]{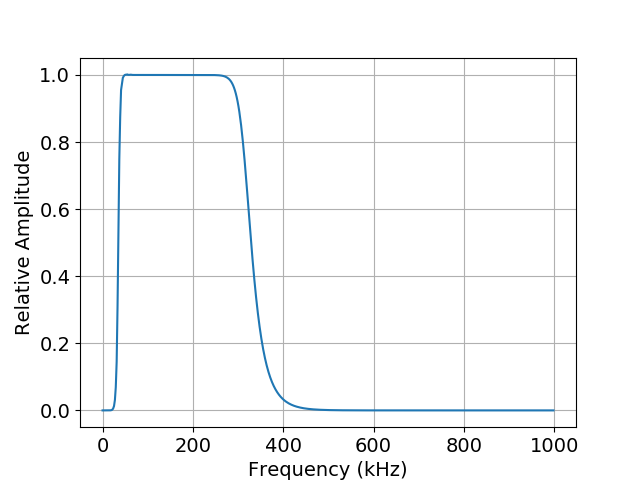}
    \captionsetup{singlelinecheck=off}
    \caption{\centering Butterworth Band-Pass Filter Amplitude Response\hspace{\textwidth}{\textcolor{darkgray}{\small\textmd{We use the butterworth bandpass filter to eliminate frequencies that fall outside of the range of important frequency content (10--500kHz)}}}}
	\label{fig:filter}
\end{figure}

First, as shown in Figure~\ref{fig:peaks}, we see frequencies drop off after
around 500kHz. To confirm this, we look at power spectral density and find that
they indeed drop off at around 500kHz and also fall outside a lower limit of
\textasciitilde10kHz. To limit the focus to just the information within these
upper and lower bounds, we apply a Butterworth band-pass filter
(Figure~\ref{fig:filter}) to filter frequencies out of this range.
Interestingly, the filter closely resembles one of decimation due to an
observable ``flat top'' where bins are not scaled relative to one another as
would be observed in Gaussian approaches. The Butterworth filter is chosen over
other filters given that there is a quicker roll-off at the cut-off frequencies
with no rippling and thus robustly preserving frequency content compared to other
linear filters.

After applying the band-pass filter, we compute the STFT to obtain the
frequency content of our signal samples over time. We then compute the Mean
Frequency Profile (MFP) -- the mean of the absolute value of each frequency over
time -- from the STFT. By computing this, we can observe both the location of
frequency peaks (variation in amplitude across movements), as well as the width
to the frequency peak (different movements may have different velocities for
each moving component). The MFP is computed as follows:

\begin{equation}
	MFP(v) = \frac{1}{M} \sum_{m=1}^{M} |STFT(v, m)| \quad MFP(v) \in \mathbb{R}^L
\end{equation}

where M is the number of time instants of STFT and L is the number of discrete
points in the Fourier transform. In the case of fingerprinting individual
movements, we can assume a constant cadence between target robot movements
over windows of time, as they are programmed to be sampled at 2s intervals. By
averaging the frequency bins over time, it is possible that some resolution is
lost regarding movements of specific components of the robot arm. However,
given that the aim is to discriminate between robot movements themselves, rather
than parts of the arm, this information is acceptable to discard.
% A visual depiction of the MFP variations for movements can be seen in Figure~\ref{fig:mfp}.
% Interestingly, for other robots that use multiple arms and tools (i.e. surgical
% robots), this information may provide useful for more fine-grained inference.
%
% \begin{figure}[H]
% 	\centering
% 	\includegraphics[width=1.0\linewidth]{mfp}
% 	\caption{MFPs for Movement Classes}
% 	\label{fig:mfp}
% \end{figure}

\begin{figure}[H]
	\centering
	\includegraphics[width=1.0\linewidth]{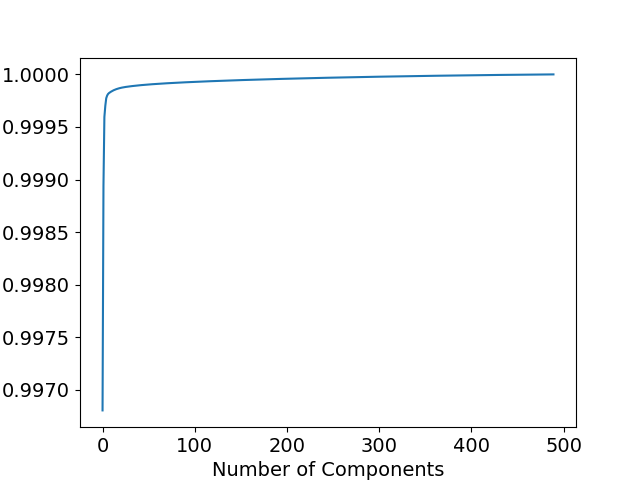}
    \captionsetup{singlelinecheck=off}
    \caption{\centering Cumulative Explained Variance for Movements\hspace{\textwidth}{\textcolor{darkgray}{\small\textmd{The cumulative explained variance shows that 14 components is enough to represent 99.999\% of the overall variation among movements}}}}
	\label{fig:pcacev}
\end{figure}

% TODO: Put in discussion: We use PCA but would other techniques (i.e. DR or approximation) such
%       				   as Nystroem approximation be useful to explore?
Even by extracting the MFP from the STFT, the resulting feature vector for each
movement sample is still fairly large and would induce a large amount of strain
on computing power for fingerprinting. To this, we apply a Principle Component
Analysis (PCA) as a dimensionality reduction technique~\cite{ma2019dimension} to decorrelate
components of the MFP to a smaller subset that still retains a high level of
discrimination among features in the feature vector. This allows most of the
information to be represented and analysed to produce the same results with
a much smaller feature vector. In Figure~\ref{fig:pcacev} we can observe the
cumulative explained variance among feature vectors for all movements -- an
accumulation of variance for each principal component. We found that 14
components is enough to represent 99.999\% of the overall variation among features.

For the last step in our feature extraction step, we apply normalisation in the
form of zero mean and standard unit variance to produce a scaled feature set to
optimise the performance in the classification stage.

Once a collection of the resulting feature vectors for our movements are put
together in our dataset, we conduct movement classification (fingerprinting) using
a C-Support Vector SVM~\cite{novakovic2011c,jian2016new} classifier. An SVM was
used for this attack, as other approaches such as deep neural networks require a
larger sample set and one of the goals of this attack is to conduct it stealthily.
Furthermore, the use of an SVM allows for more efficient computation of the movement
fingerprints and are easier to train given small datasets. The hyperparameters for
the SVM were selected using Grid Search Cross-Validation, which prompted the use
of a linear kernel with most optimal kernel parameters of $\gamma=1.0e-4$ and
$C=1.0e5$. In this work, for each movement, we have a minimum of 100 samples for
each parameter, to which we evaluate whether fingerprinting can be successful with
a small sample size and quick sample collection phase for the adversary. In the
case of larger sample sizes, other approaches such as SVM trained with Stochastic
Gradient Descent (SGD) may be more desirable in terms of computational efficiency~\cite{menon2009large}.

\begin{table*}[h]
\centering
\begin{tabular}{|c|ccc|ccc|ccc|}
	\toprule
	& \multicolumn{3}{c|}{\textbf{Blackman}} & \multicolumn{3}{c|}{\textbf{Hamming}} & \multicolumn{3}{c|}{\textbf{Hann}} \\
	& 8192 & 16384 & 32768 & 8192 & 16384 & 32768 & 8192 & 16384 & 32768  \\
	\midrule
	\textbf{Accuracy} & 93\% & 93\% & 94\% & 92\% & 95\% & 89\% & 94\% & 96\% & 95\% \\
	\midrule
	\textbf{Time (ms)} & 507 & 581 & 740 & 506 & 563 & 646 & 524 & 565 & 655 \\
	\bottomrule
\end{tabular}
\captionsetup{singlelinecheck=off}
\caption{\centering Comparison of STFT Parameters\hspace{\textwidth}{\textcolor{darkgray}{\small\textmd{We observe that the Hann window with FFT length of 16384 is the most optimal in terms of accuracy but also efficiency in computing the movement fingerprint}}}}
\label{table:stftparams}
\end{table*}

\subsubsection{Choice of STFT Parameters}

Before an evaluation of our attack, the first step is to evaluate a choice of
parameters used in the STFT step of the feature extraction process. Specifically,
we look at the FFT length and STFT windowing function. Given that our RF
samples are recorded at 2MHz sampling rate, we have a signal recorded at
2,000,000 samples per second. According to the Nyquist-Shannon Sampling
Theorem~\cite{por2019nyquist}, this means our signals can contain frequency
content up to 1MHz. Given that the STFT provides time-localised frequency
content, there is to be a trade-off between temporal and frequency resolution.
Simply, a narrow window results in better temporal resolution but poorer
frequency resolution, and vice-versa. Given that we are attempting to both
discriminate between what movements are being carried out, as well as full
operational workflows, it is important to ensure a balance between both
temporal and frequency resolutions to allow for success in both cases. Aside
from the accuracy as a choice of this first parameter, another consideration
is the time taken to compute it. As with most signal processing applications,
a relatively fast computation time is desirable, and inherently a longer FFT
takes more time to compute. Finally, the last determining factor for the choice
of STFT is the windowing function. The use of the (sliding) window function
allows for the overlapping of disjointed parts of the input signal. This aims
to decrease the amount of spectral leakage and minimise effects such as rippling,
by determining the amplitude of side lobes to distribute spectral leakage. A
typical window suitable to many applications is the Hann window, which
is a form of generalised cosine window, with other popular choices including
Hamming and Blackman windows~\cite{podder2014comparative}. In this work, we
evaluate 3 generalised cosine windows -- specifically, the Hann, Hamming and
Blackman windows -- due to demonstrable success in pulse
shaping/filtering~\cite{vigil1993application} and many other applications.
Interestingly, given a pre-computed window (as is possible with our relatively
constant cadence in movement samples), the window function should not have
any impact on computation time, but we evaluate this for completeness.

% we chose Hann 16384

In Table~\ref{table:stftparams}, we can observe the accuracy of a baseline set
of results (distance of 1 and lowest speed of 12.5mm/s) with various window
functions, FFT lengths and the time taken to compute the STFT over 100 runs.
With respect to the overall accuracy corresponding to window choice, we found
that the Hann and Hamming windows both outperformed the Blackman window, with
the Hann window having slightly higher accuracy overall. Notably, the accuracy
increases as the FFT length increases. A higher FFT length results in a higher
spectral resolution but takes longer to compute. Because of this, a key
consideration is the computation time for movement fingerprints. With respect to
the computation time, we found that the most reasonable window and FFT length
is a Hann window with an FFT length of $16384$. While the window function has no
measurable impact on computation time assuming a pre-computed window, as in our
case, we account for erroneous times due to measurement noise. In this case,
more bespoke devices such as an FPGA may provide better accuracy for measuring
computation time. In either case, with regard to FFT length, a longer window
requires a more expensive FFT and ultimately results in a non-linear increase of
computation time with length. For this study, a Hann window with FFT length of
16384 is the best choice, as a compromise of 2ms computation time is reasonable
for the accuracy increase in comparison with the same FFT length and a Hamming
window.

\subsection{Attack Results}

As per our research questions in Section~\ref{sec:goals}, our evaluation of the
results for our proposed attack will be in this order.

\subsubsection{Individual Movement Fingerprints}

As per our first research question, we initially investigate whether the
proposed adversary can fingerprint individual robot movements via unintentional
RF emissions. Simply, this will formulate a baseline set of results, consisting
of a baseline (smallest) distance of 1mm and speed of 12.5mm/s, to which we can
compare against other parameters (i.e. movement distance). For all future
parameter comparisons, the baseline results in tables are highlighted in gray.
As shown in Table~\ref{table:distanceresults}, we observe an average accuracy of
95\% for the baseline. While most movements show relatively consistent accuracy
of at least 94\%, among them we found that the YZ movement was the lowest at
91\% with a small set of samples mistaken for XYZ. Interestingly, the Z movement
shows the most success in terms of fingerprinting. Ultimately, it is clear that
Y-involved movements show the lowest accuracy. This may be due to the fact that
the base motor that handles the Y axis is also situated alongside the microprocessor,
such that the mixing of similar frequency ranges may interfere with one another
causing a lack in variation.

\begin{table}[h]
\centering
\begin{tabular}{|c|>{\columncolor{gray!30}}cccccc|}
	\toprule
	& \multicolumn{6}{c|}{\textbf{Distance (mm)}} \\
	\textbf{Movement} & 1 & 2 & 5 & 10 & 25 & 50 \\
	\midrule
	\textbf{X} & 100\% & 100\% & 100\% & 89\% & 100\% & 100\% \\
	\textbf{Y} & 93\% & 81\% & 89\% & 73\% & 100\% & 96\% \\
	\textbf{Z} & 100\% & 100\% & 93\% & 93\% & 100\% & 100\% \\
	\textbf{XY} & 89\% & 100\% & 67\% & 100\% & 100\% & 100\% \\
	\textbf{XZ} & 100\% & 88\% & 100\% & 100\% & 100\% & 92\% \\
	\textbf{YZ} & 95\% & 80\% & 50\% & 85\% & 88\% & 92\% \\
	\textbf{XYZ} & 94\% & 88\% & 62\% & 87\% & 88\% & 96\% \\
	\midrule
	\textbf{Accuracy} & 96\% & 91\% & 78\% & 89\% & 96\% & 96\% \\
	\bottomrule
\end{tabular}
\captionsetup{singlelinecheck=off}
\caption{\centering Impact of Movement Distance on Classification Accuracy\hspace{\textwidth}{\textcolor{darkgray}{\small\textmd{Movement distance provides more fine-grained information leakage and can be fingerprinted with similar accuracy to the baseline. Smaller distances have reduced accuracy likely due to similar levels of unintentional RF emissions}}}}
\label{table:distanceresults}
\end{table}

\subsubsection{Impact of Movement Distance on Fingerprinting}
The second objective concerns a higher level of granularity for movement
fingerprints. Specifically, can an adversary infer how far or how fast a
movement is being carried out? The first of these two parameters that we
evaluate is the distance of movement and how it impacts classification accuracy.
The results of this parameter can be seen in Table~\ref{table:distanceresults}.
An overview of these results suggest that fingerprinting is more successful on
average as the distance of movement increases, with the exception of the YZ
movement. At 2 distance units, we can observe a decrease in precision for most
movements, with the exception of Z, XY and XYZ movements. This is potentially
due to lowered variation among principle components between 1 and 2 distance
units. At 5 distance units, we see much better precision compared to 1 or 2
distance units, except for the YZ movement which decreases by 3\%.
Interestingly, the XY movement also decreases again compared to 2 distance units
but to similar precision as the baseline. The YZ movement in this case is
incorrectly predicted as other Y-involved movements, with most being the XYZ
movement. This same pattern is also observed for the XY movement. At 10 distance
units, we observe perfect precision for all movements, with the exception of
YZ which has no change compared to 5 units. At 25 distance units, we observe
similar precision as with 10 units, however the precision reduces for the XZ and
XYZ movements. The YZ movement increases only slightly. Finally, at 50 distance
units, the precision increases back to 100\% for XZ and increases slightly for
the XYZ movement compared to 25 units. Overall, we found that while distance
does on average increase the accuracy of movement fingerprinting, larger
distances show better success. % TODO: Why is this the case?

\begin{table}[h]
\centering
\begin{tabular}{|c|>{\columncolor{gray!30}}ccccc|}
	\toprule
	& \multicolumn{5}{c|}{\textbf{Speed (mm/s)}} \\
	\textbf{Movement} & 12.5 & 25 & 50 & 75 & 100 \\
	\midrule
	\textbf{X} & 100\% & 100\% & 100\% & 100\% & 96\% \\
	\textbf{Y} & 93\% & 83\% & 58\% & 58\% & 54\% \\
	\textbf{Z} & 100\% & 92\% & 100\% & 97\% & 100\% \\
	\textbf{XY} & 89\% & 70\% & 47\% & 68\% & 63\% \\
	\textbf{XZ} & 100\% & 100\% & 97\% & 97\% & 97\% \\
	\textbf{YZ} & 95\% & 58\% & 76\% & 55\% & 57\% \\
	\textbf{XYZ} & 94\% & 69\% & 70\% & 70\% & 45\% \\
	\midrule
	\textbf{Accuracy} & 96\% & 84\% & 85\% & 80\% & 77\% \\
	\bottomrule
\end{tabular}
\captionsetup{singlelinecheck=off}
\caption{\centering Impact of Movement Speed on Classification Accuracy\hspace{\textwidth}{\textcolor{darkgray}{\small\textmd{Movement speed provides lowered accuracy as the speed increases for movement fingerprints, particularly among Y-based movements, which may be linked to the physical infrastructure}}}}
\label{table:speedresults}
\end{table}

\subsubsection{Impact of Movement Speed on Fingerprinting}
As well as movement distance, the speed at which the movement is being carried
out may also provide a higher level of granularity to movement fingerprints. The
results for the speed parameter can be seen in Table~\ref{table:speedresults}.
Overall, it is clear that as the speed of movement increases, the variation in
the RF feature set reduces resulting in lowered classification accuracy. We
observe better fingerprinting accuracy for the XZ movement in the speed
parameter compared to the distance parameter. Taking all movements into account,
the precision for the Y-involved movements are among the poorest as the speed
increases, with many incorrectly predicted as either Y or YZ movements perhaps
due to a lack in variation between them. This may be due to the design of the
uARM robot in which Y movements making primary use of the base motor (with the
X and Z axes primarily using the right and left motors as their respective
primary motors). The distance parameter is more accurately fingerprintable than
the speed parameter, and thus would be more useful to track. While speed may be
a useful candidate still, keeping track of the distance is a more reliable
metric in most cases. For example, the speed at which an operator performs an
operation can vary depending on the context or object being moved, but the
distance may be relatively consistent for these specific avenues. In an
industrial setting, for fingerprinting alone speed may not be as useful, however
in the case of an audit in which one might wish to determine whether the robot
was behaving in an erratic fashion, speed data might prove useful through
continuous monitoring.

\begin{table}[h]
\centering
\begin{tabular}{|l|ccc|} %c
\toprule
& \multicolumn{3}{c|}{\textbf{Recovery Rate}} \\ % \textbf{Avg. Duration} & \textbf{Pos Changes} \\
\textbf{Operation} & 1 & 2 & 3 \\
\midrule
Push & 100\% & 100\% & 100\% \\
Pull & 100\% & 100\% & 100\% \\
Pick-and-Place & 100\% & 100\% & 100\% \\
Packing & 100\% & 57\% & 57\% \\
\midrule
\textbf{Accuracy} & 100\% & 88\% & 88\% \\
\bottomrule
\end{tabular}
\captionsetup{singlelinecheck=off}
\caption{\centering Workflow Reconstruction Results\hspace{\textwidth}{\textcolor{darkgray}{\small\textmd{Common warehousing workflows can be reconstructed in their entirety with much higher accuracy, compared to an approach which involves pattern-matching using individual movement fingerprints}}}}
\label{table:rfworkflowrecovery}
\end{table}

\subsubsection{Workflow Reconstruction}
Aside from exploring the efficacy of the attack on individual and combinations
of movements, which can be used in pattern-based reconstruction, it is also
interesting to determine whether higher-level warehousing workflows can be
reconstructed via the RF side channel. For this experiment, the second dataset
containing warehousing workflow samples was used, which is detailed in
Section~\ref{sec:dataset}. The results for the experiment on reconstructing
warehousing workflows can be seen in Table~\ref{table:rfworkflowrecovery}. In
this experiment, three different sets of each workflow were captured and
analysed using the same attack strategy as individual movements. Interestingly,
looking at the cumulative explained variance of the principal components for
workflows, a very similar amount of variance as with individual movements can
also be captured for entire workflows with the same number of principal
components, and thus the attack remains the same. Using the same SVM parameters,
we found that the first set of workflows achieves perfect reconstruction
accuracy. While this is a notable result, the second and third sets of these
workflows achieves similar results, with the packing operation being the
exception with 57\% accuracy in both cases. In the case of the second set, some
of the pull operations are incorrectly predicted as packing, and for the third
set, some samples of pick and place are incorrectly predicted as packing. This
may be due to the fact that the packing operation may have similarities to these
movements in the sets explored resulting in the lowered accuracy. Overall, it
is clear that using this attack approach, an adversary can very successfully
reconstruct entire operational workflows. Through the use of continuous
monitoring, entire daily operations and their performance (i.e. by also
capturing timing information) can be leaked to competitors for a potentially
malicious advantage.

% Discussion
\section{Discussion}
\label{sec:discussion}

Our proposed side channel attack using radio frequency (RF) demonstrates a
feasible approach for a insider attacker, such as a malicious
technician/operator, to place an easily disguised and economical antenna on the
factory floor. By doing so, they can fingerprint robot movements and reconstruct
entire warehousing workflows with very high accuracy. Furthermore, our
evaluation shows that such an attacker can also infer more fine-grained
information such as the speed or distance in which a movement is being carried
out.

\subsection{Defences}

Given that the radio frequencies that are captured in this attack are emitted
from an unintentional radiator source, in this case the robot, the radiated
fields require a form of mitigation (suppression) to prevent information
leakage. To recap, the robot used in this study is likely to emit unintentional
RF from its microprocessor and stepper motors, with the majority from the
microprocessor. A best defence here comes in the form of physical layer security
measures such as shielding critical portions of the microprocessor layout or
robot enclosure. Shielding against electromagnetic fields (EMF) such as RF makes
use of a barrier made of conductive or magnetic materials to isolate minimise
interference but also act as a sort of Faraday cage. The materials typically
used include copper, silver or brass, with copper being the most common for RF
shielding. In the case of our robot used, the enclosure is made of aluminium.
While this is not as conductive as copper (\textasciitilde60\%), it is usually
a second choice due to other properties such as electrical conductivity,
strength-to-weight ratio, cost and malleability. With respect to the enclosure,
a suitable defence strategy to explore is to evaluate whether a thicker
aluminium casing would provide better protection against the proposed RF side
channel attack. Furthermore, it would also be interesting to observe whether
other casing materials such as copper or brass might perform with their innate
contrasts in protective properties (i.e. conductance).

\subsection{Impact}

While we demonstrate the impact of this attack from a business perspective, in
the form of operational compromise of robotic workflows that could leak
sensitive business information, it is important to look at the impact of our
attack in other areas such as government/international specification,
regulations and international standards.

The first impact we look at is government specification. In the United States
of America, the government and NATO specification TEMPEST is used to cover
methods for eavesdropping on and protecting (shielding) against information
leakages from unintentional emanations such as RF or acoustic. While many
specifics of TEMPEST are classified, in the public domain three levels of
protection requirements are set out: NATO SDIP-27 LEVEL A, B and C. Level A
is the strictest standard which assumes the attacker has almost immediate access
to the environment or devices. Level B assumes the attacker cannot be within
20 metres and is more relaxed, and Level C assumes a distance of 100 metres.
Unfortunately, TEMPEST mainly addresses nation-state level equipment and
facilities. The guidance states that in general, devices such as these robots are
typically not qualified under TEMPEST as most of these devices, including
commercial off-the-shelf components of robots which are assumed to conform to
Level C without any modification. Further, it is not clear on specific key
requirements of shielding against unintentional emanations. Given the impact on
businesses in any respect, their confidentiality is key -- particularly against
those which they are in conflict (competition) with -- and thus the potential
of insider threat should ultimately be a cause to improve the standards of
robot protection in non-military settings.

The next impact of this attack pertains to regulation and regulatory compliance.
Guidance from regulators such as Ofcom in the UK and T{\"U}V issue rules for
compliance for all uses of RF, whether emanations occur under normal conditions
or from unintentional radiators such as the context of this work. However, such
guidance does not cover robotics systems but only a subset of typical components
(i.e. power input or cables). In the case of Ofcom, regulation only detail
where an EMF record is not required but does not include where robotics systems
or unintentional RF emanations apply. Clearly some modifications and additions
to existing regulation is needed to cover attacks like this on robotic systems.
As well as existing regulation, one key question that may arise is how can one
audit a robot? While continuous monitoring and audits of software and hardware
components can provide some guarantees, the risk of malware or invalid device
calibration that disrupts operational accuracy also needs attention.
Interestingly, we observed that from the high recovery rates of industrial
warehousing workflows, the use of the RF side channel can act as a defence that
can provide information to auditing procedures. By monitoring typically correct
robot workflows over time, any drift in accuracy from either a malicious or
unintentional {\em attack} vector could be recognised by the system, for example
through the use of LSTM networks and RF time-series information. This is a point
of future work.

Finally, we look at international standards. The key international standard that
applies to the impact of this attack is CISPR 11 for governing EMF emissions
from industrial, scientific or medical (ISM) equipment, among others, which can
use the ISM license free bands like 2.4 GHz. The ISM bands are defined by
international telecommunication union (ITU) radio regulations, which have a
variety of allocated ranges within the band of \textasciitilde6.76MHz--256GHz.
In the context of this attack, we find that our smaller robot falls outside of
ISM bands where such standards and regulation typically apply. However, for
larger industrial robots, an evaluation of this attack would be worth observing
to truly understand the impact in this case. Interestingly, an investigation
into the RF emissions related the size and general load/power requirements of
different robotic systems may also reveal that some may also fall out of the
ISM bands and may in future provoke a discussion on updating existing standards.

\subsection{Limitations}

One limitation of this work is the robot used. In terms of direct replicability
for a real-world industrial robot arm, our robot is much smaller than one
typically seen on the factory floor. While the attack may not provide the same
level of movement inference in this case as with our robot, a larger robot would
employ more motors that require more power to operate. Given that the motors
and integrated circuits still emit unintentional RF from the nature of their
operation~\cite{cobb2010physical}, the attack would simply require extra
exploratory analysis in terms of the frequency range of the RF emissions to
tune the parameters. As well as this, it would be interesting to perform an
exploratory analysis of unintentional RF emissions from a range of different
robots.

A second limitation one can consider is a disconnect between the tool equipped
at the end-effector and the motion being carried out. The robot arm used in this
study is equipped with an integrated pump with a maximum pressure of 33kPa and
maximum lifting weight of 1kg. The goal with the use of the tool was to
determine whether an attacker could identify the contents (by weight inference)
of packages being lifted. Unfortunately, there was no measurable unintentional
RF emissions between weights being lifted, nor to distinguish whether the pump
was on or off. For a larger robot however, a pump end-effector with a larger
lifting weight that requires more power may change this result. As well as this,
it would be interesting to see the impact of additional power requirements
for lifting weights with different tools (i.e. grippers) has on RF emissions.
One potential solution to this limitation in the case of our robot could be
a different side channel. For example, the acoustic side channel may be useful
with measuring the acoustic emanations for lifting heavier weights. Given that
heavier weights require more force to lift, and ultimately more power, there
may also be more sound emitted as the object is being lifted.

% Related Work
\section{Related Work}
\label{sec:related}

While there have not been any side channel literature pertaining directly to robotics systems, the use of side channels have shown success relating to
components of robotics systems, as well as similar architectures such as 3D printers. In this section we present related work on RF side channels and other
side channels and techniques related to systems that share common characteristics with robotic systems.

There are many devices which emanate unintentional RF, including microprocessors and motors, among others. Graham et al.~\cite{graham2018block} demonstrate that
RF emanations can be identified by correlating RF emissions with bit flips that produce detectable electrical pulses. This is analogous to the switching
activities in transistors as pointed out by Cobb et al.~\cite{cobb2010physical}.
In similar work, deep learning approaches have shown success operating on raw waveforms~\cite{sayakkara2019leveraging,oord2016wavenet}, such as the use of convolutional neural networks operating on time series
data~\cite{cui2016multi,bai2018empirical} or residual neural
networks~\cite{wilt2020toward}, to fingerprint IoT devices and processes running
on them. However, while these approaches are successful, our attack requires a
much smaller sample set to show similar accuracy.

Aside from the radio-frequency side channel, other side channels such as power
and acoustics also show some success. For example, Sami et al.~\cite{sami2020spying}
describe an attack via the acoustic side channel that can extract sound traces
from the vibrations reflected to lidar sensors. Several authors~\cite{chhetri2017confidentiality,acousticattackmanufacturing}
propose an acoustic side channel attack on 3D printers wherein acoustic
emanations are used to reconstruct G-code used by 3D printers which may
correspond to potentially confidential (patented) designs. Related to this,
Song et al.~\cite{song2016my} also describe a similar end goal but enhancing the acoustic
side channel using the magnetic side channel by exploiting the conductivity of
a stepper motor. Compared to our work, the extraction of G-code corresponds to
Arduino-based 3D printers. Given that our robot is also operated by an Arduino,
we show that our attack focuses solely on the movement of the robot arm and
thus reconstruction of G-code to then compromise operational confidentiality
is an unnecessary extra step. While earlier approaches make use of regression
models, more recent work make use of neural networks that require a large
labelled sample set to which our approach shows better opportunities for an
insider attacker. As well as this, given that individual movements can also be
reconstructed, pattern matching individual or permutations of these movements
could also lead to the leakage of confidential intellectual property.

% Conclusion
\section{Conclusion}
\label{sec:conclusion}

In conclusion, we showcase a novel attack on robotic systems that can leak information about confidential warehousing (industrial) workflows via a radio
frequency side channel. Instead of a deep learning approach, our attack requires a much smaller sample set and demonstrates fingerprinting accuracy, of individual robot movements as well as entire workflows, similar to attacks using neural networks. Furthermore, we show that even more fine-grained movements can be inferred from the RF side channel, such as distinguishing the distance or speed a movement is carried out with.

\begin{acks}
%\section{Acknowledgements}
	The authors are grateful for the support by the Engineering and Physical Sciences
	Research Council (11288S170484-102) and the support of the
	National Measurement System of the UK Department of Business, Energy \&
	Industrial Strategy, which funded this work as part of NPL's Data Science program.
\end{acks}

% References
\bibliographystyle{ACM-Reference-Format}
\bibliography{references_rf}

\end{document}